\documentclass[12pt,preprint]{aastex}
\shorttitle{ABSORPTION OF GAMMA RAYS IN 3C 279}

\shortauthors{Bai et al.}

\begin{document}

\title{ABSORPTION OF 10 GeV--1 TeV GAMMA RAYS FROM 3C 279}

\author{J. M. Bai\altaffilmark{1}, H. T. Liu\altaffilmark{1}, and L. Ma\altaffilmark{2}}

\altaffiltext{1}{National Astronomical Observatories/Yunnan
Astronomical Observatory, Chinese Academy of Sciences, Kunming,
Yunnan 650011, China; liuhongtao1111@hotmail.com;
baijinming@ynao.ac.cn}

\altaffiltext{2} {Physics Department, Yunnan Normal University,
Kunming 650092, China; send offprint requests to
liuhongtao1111@hotmail.com}

\begin{abstract}

In this paper, we revisit gamma-ray--emitting region for 10 GeV--1
TeV gamma rays from 3C 279 through studying the photon-photon
absorption optical depth due to the diffuse radiation of the
broad-line region (BLR) and the extragalactic background light
(EBL). Based on the power-law spectrum detected by MAGIC, the
preabsorbed spectra are inferred by correcting the photon-photon
absorption on the diffuse photons of the BLR (internal absorption)
and the EBL (external absorption). Position of gamma-ray emitting
region $R_{\rm{\gamma}}$ determines the relative contributions of
this two diffuse radiation to the total absorption. Our results
indicate that $R_{\rm{\gamma}}$ may be within the BLR shell for 3C
279, likely closer to the inner radius, which is consistent with
our previous results. This is neither consistent with the
suggestions of B\"ottcher et al. (2008b), that VHE gamma-ray
emission is produced far outside the BLR, nor with the assumptions
of Tavecchio \& Mazin (2008), that VHE gamma-ray--emitting region
is inside the BLR cavity. $R_{\rm{\gamma}}$ is a key physical
quantity that could set some constraints on emission mechanisms
that produce the VHE gamma rays from 3C 279. Observations of $\it
Fermi$-LAT, MAGIC, HESS, and VERITAS in the near future could give
more constraints on the position of gamma-ray emitting region
relative to the BLR.

\end{abstract}

\keywords{gamma rays: theory --- quasars: individual (3C 279)}

\section{INTRODUCTION}
The classical flat spectrum radio quasar (FSRQ) 3C 279 is one of
the brightest extragalactic objects in the gamma-ray sky. It lies
at a redshift of $z=0.536$ (Marziani et al. 1996). It was detected
by EGRET, and its spectrum does not show any signature of
gamma-ray absorption by pair production up to $\sim$10 GeV
(Fichtel et al. 1994; von Montigny et al. 1995). With the
detection of high energy gamma rays from 66 blazars, including 51
FSRQs and 15 BL Lacertae (BL Lac) objects, in the GeV energy range
by EGRET (Fichtel et al. 1994; Thompson et al. 1995, 1996;
Catanese et al. 1997; Lin et al. 1997; Mukherjee et al. 1997;
Villata et al. 1997; Hartman et al. 1999; Nolan et al. 2003), an
exceptional opportunity is presented for understanding the central
engine operating in blazars. Venters \& Pavlidou (2007) found that
the most likely Gaussian intrinsic spectral index distribution for
EGRET blazars has a mean of 2.27 and a standard deviation of 0.20,
as well as some indication that FSRQs and BL Lac objects may have
different intrinsic spectral index distributions (with BL Lac
objects being harder). Bloom (2008) confirmed the radio and
gamma-ray correlation of EGRET blazars, and replicated through
Monte Carlo simulations the observed luminosity relationship if a
synchrotron self-Compton model is assumed. Recently, Thompson
(2008) summarized results from EGRET and reported that a central
feature of the EGRET results was the high degree of variability
seen in many gamma-ray sources, indicative of the powerful forces
at work in objects visible to gamma-ray telescopes. The first
three months of sky-survey operation with the Fermi Gamma Ray
Space Telescope ($Fermi$) Large Area Telescope (LAT) revealed 132
bright sources (Abdo et al. 2009). Associations of 106 of these
sources are indicated on high-confidence with known AGNs: two
radio galaxies, namely Cen A and NGC 1275, and 104 blazars
consisting of 57 FSRQs, 42 BL Lac objects, and 5 blazars with
uncertain classification (Abdo et al. 2009). Four new blazars were
discovered on the basis of the LAT detections, and only thirty
three of the 106 sources were previously detected with EGRET (Abdo
et al. 2009). It was revealed for the LAT blazars that the average
GeV spectra of BL Lac objects are significantly harder than the
spectra of FSRQs (Abdo et al. 2009). Some BL Lac objects have also
been firmly detected by atmospheric Cerenkov telescopes at
energies above 1 TeV, such as Mrk 421 (Punch et al. 1992), and Mrk
501 (Quinn et al. 1996). At present, 27 active galactic nuclei
(AGNs) have been detected in very high energy (VHE) gamma rays,
including 23 BL Lac objects, two radio sources (M 87 and Cen A),
an unidentified MAGIC source (MAGIC J0223+430, possibly associated
with the radio galaxy 3C 66B), and the first FSRQ, 3C 279, which
has the highest redshift among these VHE
AGNs\footnote{http://www.mppmu.mpg.de/$\sim$rwagner/sources/}.
Anderhub et al. (2008) reported upper limits to the VHE flux of
the FSRQ 3C 454.3 ($z=0.859$) derived by the Cherenkov telescope
MAGIC during the high states of July/August and November/December
2007. The positive detection of 3C 279 in VHE gamma rays by the
MAGIC telescope (Albert et al. 2008) comes unexpectedly. It was
suggested that FSRQs are unlikely to be important VHE gamma-ray
emitter, but BL Lac objects are important emitter (e.g., Fossati
et al.1998; Ghisellini et al. 1998; B\"ottcher \& Dermer 2002).
Observations showed that these suggestions are right before 3C 279
was detected by MAGIC. The most recent calculations showed that
the internal absorption could significantly annihilate VHE gamma
rays from FSRQs (Donea \& Protheroe 2003; Liu \& Bai 2006; Reimer
2007; Aharonian et al. 2008b; Liu et al. 2008; Sitarek \& Bednarek
2008; Tavecchio \& Mazin 2009).

The gamma rays from blazars are generally believed to be
attributable to emission from a relativistic jet oriented at a
small angle to the line of sight (Blandford \& Rees 1978). The
diffuse radiation field of broad-line region (BLR) could have a
strong impact on the expected external Compton (EC) spectra of the
most powerful blazars, FSRQs (e.g., Sikora et al.1994; Liu \& Bai
2006; Reimer 2007; Aharonian et al. 2008b; Liu et al. 2008;
Sitarek \& Bednarek 2008; Tavecchio \& Ghisellini 2008; Tavecchio
\& Mazin 2009). Not only do the external soft photon fields from
the BLR provide target photons for the EC processes to produce
these gamma-ray components, they also absorb gamma rays from the
EC processes by photon-photon pair production. Many efforts to
study the absorption of gamma rays have focused on photon-photon
annihilation by the diffuse extragalactic background radiation in
the IR, optical, and UV bands (e.g., Stecker et al. 1992; Stecker
\& de Jager 1998; Oh 2001; Renault et al. 2001; Chen et al. 2004;
Dwek \& Krennrich 2005; Schroedter 2005; Stecker et al. 2006;
Stecker \& Scully 2009; Tavecchio \& Mazin 2009). This external
absorption of gamma rays by the diffuse extragalactic background
light (EBL) has also been proposed to probe the EBL itself
(Renault et al. 2001; Chen et al. 2004; Dwek \& Krennrich 2005;
Schroedter 2005; Stecker \& Scully 2009; Tavecchio \& Mazin 2009).
Indeed, the absorption of gamma rays inside FSRQs could result in
serious problems for the possibility of using the external
absorption of gamma rays to probe the IR-optical-UV extragalactic
background, because it could mask the intrinsic gamma-ray spectra
(Donea \& Protheroe 2003; Liu \& Bai 2006; Reimer 2007; Liu et al.
2008). The intrinsic spectra of gamma rays are complicated by the
complex spectrum of the diffuse radiation field of the BLR in
FSRQs (Tavecchio \& Ghisellini 2008; Tavecchio \& Mazin 2009).

The position of the gamma-ray--emitting region is still an open
and controversial issue in the researches on blazars (Ghisellini
\& Madau 1996; Georganopoulos et al. 2001; Lindfors et al. 2005;
Sokolov \& Marscher 2005). In our previous work (Liu \& Bai 2006,
hereafter Paper I; Liu et al. 2008, hereafter Paper II), the
position of the gamma-ray--emitting region was a key parameter to
determine whether gamma rays could escape the diffuse radiation
field of the BLR for FSRQs. VHE gamma rays will be strongly
attenuated for 3C 279 if the emitting region is within the BLR
cavity (see Paper I \& II). However, MAGIC detected VHE gamma-ray
spectrum from 3C 279 (Albert et al. 2008). H.E.S.S. observations
placed an upper limit on the integrated photon flux for 3C 279
(Aharonian et al. 2008a). In Paper I, we addressed an important
topic in gamma-ray astrophysics, namely, the absorption of
high-energy gamma rays inside FSRQs by photons of the BLR. In
Paper II, we addressed the particular topic of absorption in the
gamma-ray quasar 3C 279 using the available observational data,
the integrated photon flux measured by MAGIC (Teshima et al. 2007)
and the upper limit placed by H.E.S.S (Aharonian et al. 2008a),
and its potential effect on the gamma-ray spectrum. Photon index
of $\gtrsim 6.4$ was limited and used to constrain the position of
gamma-ray--emitting region in 3C 279 (see Paper II). In this
paper, we attempt to address the same topic as in Paper II by
using the VHE gamma-ray spectrum detected by MAGIC.

The structure of this paper is as follows. $\S$ 2 presents
intensity of VHE gamma rays. $\S$ 3 presents theoretical
calculation of photon-photon optical depth for 3C 279. $\S$ 4
presents spectral shape of VHE gamma rays. $\S$ 5 is for
discussions and conclusions. Throughout this paper, we use a flat
cosmology with a deceleration factor $q_0=0.5$ and a Hubble
constant $H_0=75 \/\ \rm{km \/\ s^{-1} \/\ Mpc^{-1}}$.

\section{INTENSITY OF VHE GAMMA RAYS}
The classical FSRQ 3C 279 is one of the brightest extragalactic
objects in the gamma-ray sky, and it is also the first VHE
gamma-ray FSRQ. on the night of 2006 February 23, the observations
showed a clear gamma-ray signal with an integrated photon flux
$F(E_{\gamma}>200\/\ \rm{GeV})=(3.5\pm0.8)\times 10^{-11}\/\
\rm{cm^{-2}s^{-1}}$ (Teshima et al. 2007). H.E.S.S. observations
in 2007 January measured an upper limit of integrated photon flux
$F(E_{\gamma}>300\/\ \rm{GeV})<3.98\times 10^{-12}\/\
\rm{cm^{-2}s^{-1}}$ (Aharonian et al. 2008a). VHE gamma-ray
emission from 2006 February 22--23 is likely due to an
intermediate state, and on the rest of the nights it likely
correspond to a low state (Paper II). GeV gamma-ray emission from
2000 February 8 to March 1 and VHE gamma-ray emission from 2006
February 22--23 likely originated from very similar states (see
Paper II).

Recently, VHE spectrum of 3C 279 has been published from the MAGIC
collaboration (Albert et al. 2008), and the photon intensity of
the VHE gamma rays is
\begin{equation}
\frac{dI}{dE_{\gamma}}=(5.2\pm1.7)\times 10^{-13}
\left(\frac{E_{\gamma}}{200\rm{GeV}}\right)^{-(4.11\pm0.68)}\/\
\rm{cm^{-2}s^{-1}GeV^{-1}}.
\end{equation}
In Paper II, spectral index of $\gtrsim 6.4$ is limited by the
integrated photon fluxes measured by MAGIC and HESS, because no
VHE spectrum of 3C 279 has been published before Paper II. Within
error, the spectral slope of $4.11$ is consistent with the maximum
spectral slope of $4.21$ measured in the VHE regime for PG
1553+113 (Albert et al. 2007). The value of $4.11$ is larger than
the slope of $2.3$ for the VHE spectrum extrapolated from the GeV
spectrum observed on 2000 February 8 to March 1 (Hartman et al.
2001). The average EGRET blazar spectrum was found to have a slope
of $2.27$ (Venters \& Pavlidou 2007). The mean of the intrinsic
spectral indices is $\overline{\Gamma}\approx 2.3$ for 17 blazars
detected in the VHE regime (see Wagner 2008). Stecker et al.
(1992) investigated the photon-photon absorption of the VHE
gamma-ray spectrum, extrapolated from the differential spectrum of
gamma rays measured by EGRET during 1991 June, on the
extragalactic background infrared radiation field. The corrected
photon flux is not inconsistent with the upper limit from the
Whipple observatory (Stecker et al. 1992). This indicates that the
VHE and GeV gamma rays likely have some relationship.

\section{Photon-Photon Optical Depth}
The internal absorption of gamma rays due to the diffuse photons
of the BLR is adopted from Paper II for 3C 279 (see Figs. 2, 3,
and 4). Stecker et al. (1992) first pointed out the importance of
the EBL in determining the opacity of the universe to high energy
gamma rays at higher redshifts, and investigated the infrared EBL
absorption on high energy gamma rays for 3C 279 and found an
absorption optical depth of $3.7 \lesssim \tau \lesssim 9.7$ for 1
TeV gamma rays. Dwek \& Krennrich (2005) detailed the
observational limits and detections of the EBL, and the relevant
EBL spectral templates. Their investigation showed that the
absorption of gamma rays at $\lesssim$1 TeV is entirely
contributed by the EBL at 0.1--10 $\rm{\mu m}$ (see Fig. 3 of Dwek
\& Krennrich 2005). They only calculated the optical depth for low
redshift sources. Stecker et al. (2006) have given the optical
depth for sources at redshifts z$<$6.  An analytic form to
approximate the function $\tau_{\gamma\gamma}(E_{\gamma},z)$ of
the external EBL absorption is given in Stecker et al. (2006).
Five models are considered in Stecker \& Scully (2009) to
calculate the EBL absorption for 3C 279, and the five EBL
absorptions were approximated as
\begin{equation}
\log \tau_{\gamma\gamma}=a_1 x^3+a_2 x^2+a_3 x+a_4,
\end{equation}
where $x\equiv\log E_{\gamma}(\rm{GeV})$. Coefficients $a_1$
through $a_4$ are given in Table 1 (Stecker \& Scully 2009). We
adopt these coefficients for five EBL models to calculate the EBL
absorption optical depth.

In order to compare the internal and external absorption, the two
absorptions are presented in Figures 1$a$, 2$a$, and 3$a$. The
external absorption monotonically increases with gamma-ray energy.
If $R_{\rm{\gamma}}$ is around the inner radius $r_{\rm{BLR,in}}$,
the internal absorption dominates over the external absorption,
and the latter mainly presents itself in the VHE interval and is
much less than unity from around 10 to a few tens of GeV. If
$R_{\rm{\gamma}}$ is around the median of $r_{\rm{BLR,in}}$ and
$r_{\rm{BLR,out}}$, the total absorption is dominated by internal
absorption in the 10--100 GeV interval and the external absorption
dominates from around 400 GeV to 1 TeV (see Fig. 2$a$). The
relative contributions of the internal and external absorption to
the total are comparable from around 100 to 400 GeV (see Fig.
2$a$). If $R_{\rm{\gamma}}$ is around the outer radius
$r_{\rm{BLR,out}}$, the internal absorption is comparable to the
external around 10 GeV, and the former is dominated by the latter
from around 30 GeV to 1 TeV (see Fig. 3$a$). The position of
gamma-ray--emitting region determines the relative contributions
of the internal and external absorption to the total photon-photon
annihilation optical depth for 10 GeV--1 TeV gamma rays.

\section{SPECTRAL SHAPE OF VHE GAMMA RAYS}
Measured and intrinsic VHE gamma-ray spectra can be well described
by a power-law, and the intrinsic gamma-ray spectra of 17 BL Lac
objects have been inferred by only correcting for the EBL absorption
in the measured spectra (Wagner 2008). If the gamma-ray--emitting
region is far from the BLR in FSRQs, it is reasonable to infer the
intrinsic spectra by only correcting for the EBL absorption.
Otherwise, this approach is likely to be insufficient for FSRQs to
infer the intrinsic spectra from the measured VHE spectra, because
the internal absorption is not negligible when compared with the
external absorption (see Figs. 1$a$ and 2$a$). The position of
gamma-ray--emitting region determines the relative contributions of
the internal and external absorption. The dependence of the internal
absorption on gamma-ray energy relies on the position of the
emitting region (see Figs. 1$a$, 2$a$, and 3$a$). The external
absorption monotonically increases with gamma-ray energy, and thus
its effect on the shape of spectrum is more straightforward than
that of the internal absorption. It is obvious that the external
absorption softens the observed gamma-ray spectrum relative to the
emitted one. After correcting the gamma-ray spectra for the internal
and external absorption, we show the preabsorbed spectra for three
values of $R_{\gamma}$ in Figures 1$b$, 2$b$, and 3$b$ (color
lines). For $R_{\gamma}$ is around $r_{\rm{BLR,in}}$, the
preabsorbed VHE gamma-ray spectra peak from around 200 to 400 GeV
(Fig. 1$b$). If $R_{\gamma}\gtrsim
(r_{\rm{BLR,in}}+r_{\rm{BLR,out}})/2$, the preabsorbed VHE gamma-ray
spectra are concave except these green lines in Figs. 2$b$ and 3$b$.
After passing through the internal and external diffuse radiation
fields, the detected gamma-ray spectra are softer than the
preabsorbed ones, when $R_{\gamma}$ range from around 0.25 to 0.4
$\rm{pc}$ (see Figs. 2$b$ and 3$b$).

These preabsorbed VHE gamma rays from 500 GeV to 1 TeV can be well
fitted by a power-law spectrum. For $R_{\gamma}=0.1 \/\ \rm{pc}$,
photon indices of $2.14\pm0.05$ to $4.95\pm0.06$ are given by the
fit for color dotted lines; values of $1.87\pm0.15$ to
$5.92\pm0.14$ are given for color dashed lines; and values of
$6.36\pm0.15$ to $7.54\pm0.15$ are given for green lines (Fig.
1$b$). The values of $6.36\pm0.15$ to $7.54\pm0.15$ for green
lines are larger than the typical value of $2.3$ for the known VHE
sources and are also larger than the intrinsic photon index of
$\simeq 3.6$ for PG 1553+113, the maximum among the known VHE
spectra (Wagner 2008). Values of $2.14\pm0.05$ to $4.95\pm0.06$
for color dotted lines and values of $1.87\pm0.15$ to
$5.92\pm0.14$ for color dashed lines are not inconsistent with the
range of intrinsic photon index from 1.3 to 3.6 for the known VHE
sources (Wagner 2008). For $R_{\gamma}=0.25 \/\ \rm{pc}$ and
$R_{\gamma}=0.4 \/\ \rm{pc}$, the preabsorbed VHE gamma-ray
spectra are concave except these described by the green lines
(Figs. 2$b$ and 3$b$). Values of $1.69\pm0.12$ to $1.81\pm0.12$
are given for green lines in Fig. 2$b$ and values of $1.08\pm0.10$
are given for green lines in Fig. 3$b$. These are smaller than the
typical value of 2.3 for the known VHE sources.

Five models considered in Stecker \& Scully (2009) to calculate
the EBL absorption for 3C 279 are the fast evolution and baseline
models of Stecker et al. (2006), the best-fit and low-IR models of
Kneiske et al. (2004), and the model of Primack et al. (2005). For
$R_{\gamma}\gtrsim 0.25 \/\ \rm{pc}$, the preabsorbed VHE
gamma-ray spectra are concave as the Stecker et al. models and the
Kneiske et al. models are used, but convex for the Primack et al.
model. After VHE gamma-ray spectra measured by MAGIC are corrected
for the EBL, the preabsorbed spectra of 3C 279 have a concave
shape (Albert et al. 2008; Tavecchio \& Mazin 2009).

\section{DISCUSSIONS AND CONCLUSIONS}

In Paper I and II, the template of the diffuse BLR radiation
includes the continuum in the IR-optical-UV band. This diffuse
continuum is presented by the diluted blackbody radiation in Paper
I, and is presented by the diluted multi-temperature blackbody
radiation in Paper II. The details of the diluted blackbody are
illustrated by equation (20) in Paper I and the first paragraph of
section 3 in Paper II. Though, these approximation to the
continuum from the BLR gas differ from the realistic continuum
emitted by the BLR gas, these approximate continua have an
important role in shaping the optical depth to gamma rays at
energies $>$ 100 GeV (see Figs. 2$a$-4$a$ in Paper II, and Figs.
1$a$-3$a$ in this paper). In Paper II and this paper, the
approximate energy spectrum of the continuous part of the BLR
emission is given by the normalized mult-temperature blackbody
spectrum integrated all over the standard accretion disk, similar
to the case E in Tavecchio \& Mazin (2009). The optical depth to
gamma rays from 10 GeV to 1 TeV is similar to that of the case E
(see red lines in Fig. 3 of Tavecchio \& Mazin 2009). The diluted
multi-temperature blackbody spectrum has a long low-energy tail in
IR-optical band, and this feature is similar to that from
simulation run by Tavecchio \& Ghisellini (2008) and Tavecchio \&
Mazin (2009). B\"ottcher et al. (2008b) confirmed the findings of
Liu et al. (2008): the dependence of the photon-photon absorption
depth on the dimensionless photon energy and the location of the
gamma-ray production site. These similarity and confirmation
indicate that our approximation to the continuum from the BLR gas
is reasonable. Position of gamma-ray emitting region
$R_{\rm{\gamma}}$ can significantly affect the shape of the
optical depth to gamma rays from 10 GeV to 1 TeV, including the
emission line and continuum absorption optical depth and the total
one (see Figs. 2$a$-4$a$ in Paper II, and Figs. 1$a$-3$a$ in this
paper). The peak of $\tau_{\gamma\gamma}(E_{\gamma})$ tends to
move towards higher energy as $R_{\rm{\gamma}}$ increases. The
integral over $R$ has a upper limit of $R_2=r_{\rm{BLR}}$ in
equation (1) in Tavecchio \& Mazin (2009), where $r_{\rm{BLR}}$ is
equivalent to $r_{\rm{BLR,in}}$ in our papers. However, this upper
limit of $R_2$ adopted in Paper I and II and this paper is much
more than $r_{\rm{BLR}}$. This difference of this upper limit
results in a steeper photon-photon absorption optical depth than
that in Tavecchio \& Mazin (2009) (see Fig. 4). The absorption due
to the soft photons outside $r_{\rm{BLR}}$ is comparable to or
larger than that from these soft photons insides $r_{\rm{BLR}}$
(see Fig. 4), and can not be ignored in calculations of the total
absorption. Furthermore, the two absorptions have different
profiles in the VHE regime due to the two different integral
ranges of $R$. These effects are presented in Figure 4. For
comparison to illustrate these effects, the lower dashed and
dash-dotted lines, corresponding to the total absorption optical
depth and the absorption optical depth due to the diffuse
continuum from the BLR, are shifted upward to the solid and
dash-double-dotted lines, respectively. These curves in Figure 4
are estimated as $a_{\rm{\ast}}=0.5$, $r_{\rm{BLR,in}}=0.1\/\
\rm{pc}$, $r_{\rm{BLR,out}}=0.4 \/\ \rm{pc}$, and
$R_{\rm{\gamma}}=2\times 10^{16} \/\ \rm{cm}$ which is the same as
in Tavecchio \& Mazin (2009). There exists an important (almost
flat) optical-IR component in the model of Tavecchio \& Mazin
(2009), and this component dominates $\tau_{\gamma\gamma}$ at
energies above 100 GeV. This component could result in a bump
around 10 TeV in $\tau_{\gamma\gamma}$. In the case E of Tavecchio
\& Mazin (2009), there is another bump around 100 GeV in
$\tau_{\gamma\gamma}$ (see red lines in the upper panel, Fig. 3).
There are also valleys between the two bumps in the case E. Due to
the steepening effect of $\tau_{\gamma\gamma}$ as the upper limit
$R_2$ of the integral over $R$ increases, these valleys become
steeper as $R_2$ increases. Thus, the case E of Tavecchio \& Mazin
(2009) could produce much more pronounced variations in the
opacity from 100 GeV to 1 TeV if $R_2>> r_{\rm{BLR}}$. We think
that the profiles of these curves presented in Figure 1 should be
reliable. Thus it is likely to see the presence of the important
bump in the intrinsic spectrum around 300 GeV in some of the cases
(e.g., Fig. 1$b$ in this paper and Fig. 2$b$ in Paper II). One
more realistic approach to get the energy spectrum emitted by the
BLR gas is as follows: on the basis of the observational energy
spectrum in IR-optical-UV band with the high resolution and the
high ratio of signal to noise, the BLR energy spectrum can be
derived by subtracting narrow emission lines, and the continua
emitted by jet and host galaxy from the observed spectrum of
blazars. The BLR spectra obtained by this way should be relatively
closer to the realistic ones than these assumed or simulated
previously.

Five EBL models are used in this paper to estimate the EBL
absorption to gamma rays. The model of Primack et al. (2005)
exhibits a steep mid-IR valley that is directly conflict with
solid lower limits obtained from galaxy counts from observations
at mid-IR wavelengths (Altieri et al. 1999; Elbaz et al. 2002),
because this model does not take the warm dust, polycyclic
aromatic hydrocarbon, and silicate emission components of mid-IR
galaxy spectra into account, but was based on strictly theoretical
galaxy spectra (e.g., Stecker \& Scully 2009). This confliction is
clearly shown in Figure 2 of the supplemental online material of
Albert et al. (2008). Even so, we still consider the model of
Primack et al. as one possible EBL model to estimate the EBL
absorption. However, this model only gives a lower limit to the
EBL absorption. The fast evolution model is favored by the
$Spitzer$ observations (e.g., Stecker et al. 2007). It provides a
better description of the deep $Spitzer$ number counts at 70 and
160 $\mu\rm{m}$ than the baseline model. However, $GALEX$
observations indicate that the evolution of UV radiation for $z<1$
may be more consistent with the baseline model, and the 24 $\mu
\rm{m}$ $Spitzer$ source counts are closer to the baseline model
than the fast evolution model (e.g., Stecker et al. 2007). Albert
et al. (2008) suggested that the detection of 3C 279 in VHE regime
would appear to disfavor the EBL models of Stecker et al. (2006).
Recently, Stecker \& Scully (2009) concluded that the five EBL
models used in this paper equally produce reasonable fits to the
observational data of 3C 279. Aharonian et al. (2006) argued that
intrinsic spectra must have spectral index of $\Gamma \geq 1.5$
for blazars. Combining this assumption with HESS observations of
1ES 1101-232, they placed an upper limit on the EBL at 1.5 $\mu
\rm{m}$. This limit is consistent with the baseline model, but not
with the fast evolution model that is favored by the $Spitzer$
observations. Stecker et al. (2007) reexamined the assumption of
spectral index of $\Gamma \geq 1.5$ made by Aharonian et al.
(2006). They found that relativistic shock acceleration can
produce particle population with a significant range of spectral
indices, including those with $\Gamma_e \le 2$ corresponding to
inverse Compton gamma-ray spectra with $\Gamma \le 1.5$.
Franceschini et al. (2008) found that the de-absorbed TeV spectra
of BL Lac objects are all softer than $\Gamma=1.5$, as assumed in
Aharonian et al. (2006). Later, Krennrich et al. (2008) showed
that the differential spectral index of intrinsic spectrum is
$\Gamma=1.28\pm0.20$ or harder for the three TeV blazars 1ES
0229+200, 1ES 1218+30.4, and 1ES 1101-232. Franceschini et al.
(2008), based on an impressive amount of observational data,
showed that the level of the EBL should be close to what is
estimated through galaxy counts. As pointed out in Krennrich et
al. (2008), their result is based on observational constraints,
but Franceschini et al. (2008) is based on theoretical modelling.
Thus it is allowable $\Gamma < 1.5$. These low values of $\Gamma <
1.5$ indicates that the upper limit on the near IR EBL obtained by
Aharionian et al. (2006) is allowed to increase to higher level.
Such a result is consistent with the fast evolution model. Thus,
the five models used in this paper, the baseline and fast
evolution models of Stecker et al. (2006), the best-fit and low-IR
models of Kneiske et al. (2004), and the model of Primack et al.
(2005), are partly favored or not fully supported by observations.
Then the five EBL models are likely to be reliable. In
consequence, the preabsorbed spectra (described by the color lines
in Figures 1$b$, 2$b$, and 3$b$) should be also reliable.
According to comparisons of intrinsic photon indices discussed in
$\S 4$, photon indices of 6.4 to 7.5 for the green lines in Figure
1$b$ are larger than the maximum among the known VHE spectra, and
photon indices of 1.1 for the green lines in Figure 3$b$ are
smaller than the minimum among the known VHE spectra. Thus
$R_{\rm{\gamma}}$ should be between the inner and outer radii of
the BLR. For the gamma-ray emitting region outside the median of
inner and outer radii of the BLR, the preabsorbed gamma-ray
spectra have concave shape in the range from several ten GeVs to
TeV (Figs. 2$b$ and 3$b$).

The intrinsic spectra of Mrk 421 and Mrk 501 were derived for all
the realizations of the EBL templates, and some EBL realizations
led to an unphysical behavior in the blazar intrinsic VHE
spectrum, characterized by an exponential rise after a decline or
flat behavior with energy (Dwek \& Krennrich 2005). These
intrinsic VHE gamma-ray spectra with physical behavior mostly show
power-law or convex shapes. Measured and intrinsic VHE gamma-ray
spectra can be well described by a power-law for 17 BL Lac objects
(Wagner 2008). Many researches on the spectrum energy distribution
(SED) of VHE gamma-ray have not shown concave shape, but convex
shape for blazars (e.g., B\"ottcher et al. 2008a, b; Dermer et al.
2009; Mannheim \& Biermann 1992; Saug\'e \& Henri 2004; Tavecchio
\& Chisellini 2008). Thus, it is reliable that the intrinsic VHE
spectra of blazars should generally have convex shapes. For
$R_{\gamma}$ beyond the median of inner and outer radii of the
BLR, the SEDs, $E_{\gamma}^2 dI/dE_{\gamma}$, of the preabsorbed
spectra denoted by the red and blue dotted lines and the red
dashed lines in Figures 2$b$ and 3$b$ have concave shapes in the
range from several ten GeVs to TeV. This indicates that
$R_{\gamma}$ is likely to be inside the median of inner and outer
radii of the BLR for 3C 279. For $R_{\gamma}=r_{\rm{BLR,in}}$, the
internal absorption optical depth is larger than unity for 10 GeV
gamma rays, and this is not consistent with observations of EGRET
(see Paper I \& II). This indicates $R_{\gamma}\gtrsim
r_{\rm{BLR,in}}$. According to comparisons of intrinsic photon
indices discussed in $\S 4$, $R_{\gamma}\sim r_{\rm{BLR,in}}$ is
allowed. Thus, the gamma-ray emitting region is likely between the
inner radius and the median of inner and outer radii of the BLR,
and is probably closer to the inner radius.

B\"ottcher et al. (2008b) studied the simultaneous optical, X-ray,
and VHE gamma-ray data from the day of the VHE detection for 3C
279, and discussed the implications of the SED for jet models of
blazars. A hadronic model is proposed to explain the SED of 3C 279
by Mannheim \& Biermann (1992). B\"ottcher et al. (2008b) showed
that the observed SED of 3C 279 can be reasonably well reproduced
by a hadronic model. However, a rather extreme jet power is
required by various versions of the hadronic model. They also
employed the homogeneous leptonic jet models, including the EC and
synchrotron self Compton (SSC) models, to explain the simultaneous
SED of 3C 279. The EC models can reproduce the simultaneous
optical and VHE gamma-ray spectrum of 3C 279, but require either a
unusually low magnetic field or an unrealistically high Doppler
factor, as well as unable to reproduce the observed X-ray data.
The SSC models can reproduce the simultaneous X-ray--VHE gamma-ray
SED of 3C 279, but fail to reproduce the optical data. The SSC
models require the gamma-ray emitting region far outside the BLR.
Bai \& Lee (2001) predicted existence of large scale synchrotron
X-ray jets in radio-loud active galactic nuclei, especially, the
X-ray jet is bright on 10 $\rm{kpc}$ scales in most red blazars
and red blazar-like radio galaxies. According to their
predictions, the large scale synchrotron X-ray jets can produce
VHE gamma rays by the SSC processes. In this case, the VHE gamma
rays from 3C 279 are produced by the same ways as in the high
peaked-frequency BL Lac objects (HBL). However, the cooling rate
in the inner jets of HBLs is lower than that of FSRQs due to the
thinner external soft photons in HBLs. Thus, electron population
in jets of HBLs can earlier gain more energies enough to emit VHE
gamma rays in a closer region to the central engines than that of
FSRQs. If the VHE gamma rays from 3C 279 are emitted by the large
scale synchrotron X-ray jet on $\rm{kpc}$ scales, the VHE
gamma-ray emitting region is likely to be imaged by $Fermi/LAT$,
and then the SSC processes could be tested and the gamma-ray
emitting position could be constrained by observations of
$Fermi/LAT$. The gamma-ray emitting position is a key physical
quantity that constrains radiation mechanisms that produce the VHE
gamma rays from 3C 279.

In this paper, we revisit the position of gamma-ray--emitting
region for 10 GeV--1 TeV gamma rays from 3C 279 through studying
the photon-photon absorption optical depth due to the diffuse
radiation of the BLR and the EBL. Based on the power-law spectrum
detected by MAGIC, the preabsorbed spectra are inferred by
correcting the internal absorption by the BLR and the external
absorption by the EBL. For the gamma-ray emitting region outside
the median of inner and outer radii of the BLR, the preabsorbed
gamma-ray spectra have concave shape in the range from several ten
GeVs to TeV (Figs. 2$b$ and 3$b$). However, many researches on
intrinsic VHE gamma-ray spectra showed power-law or convex shapes
in the VHE regime for blazars (e.g., B\"ottcher et al. 2008a, b;
Dermer et al. 2009; Mannheim \& Biermann 1992; Saug\'e \& Henri
2004; Tavecchio \& Chisellini 2008; Wagner 2008). Thus, it is
likely $R_{\rm{\gamma}}\lesssim
(r_{\rm{BLR,in}}+r_{\rm{BLR,out}})/2$ for 3C 279. Based on studies
of the photon-photon optical depth and the intrinsic spectral
indices, $R_{\rm{\gamma}}\gtrsim r_{\rm{BLR,in}}$. Thus, the
gamma-ray emitting region in 3C 279 is likely between the inner
radius and the median of inner and outer radii of the BLR,
probably closer to the inner radius. This is consistent with our
previous results (Paper II). However, this is neither consistent
with the suggestions of B\"ottcher et al. (2008b), that VHE
gamma-ray emission is produced far outside the BLR for 3C 279, nor
with the assumptions of Tavecchio \& Mazin (2009), that VHE
gamma-ray--emitting region is inside the BLR cavity for 3C 279.
This also is neither consistent with suggestions that
$R_{\rm{\gamma}}$ lies within the BLR cavity for powerful blazars
(Ghisellini \& Madau 1996; Georganopoulos et al. 2001), nor
consistent with suggestions that $R_{\rm{\gamma}}$ are outside the
BLRs for powerful blazars (Lindfors et al. 2005; Sokolov \&
Marscher 2005). $R_{\rm{\gamma}}$ determines the relative
contributions of the BLR and the EBL to the total absorption, and
is a key physical quantity that constrains emission mechanisms
that produce the VHE gamma rays from 3C 279. Observations of $\it
Fermi/LAT$, MAGIC, HESS, and VERITAS in the near future could give
more constraints on $R_{\gamma}$. Publications of intrinsic
spectra predicted by theoretical researches and these measured by
observations in the VHE regime could give stronger constraints on
the position of gamma-ray emitting region relative to the BLR.

\acknowledgements We are grateful to the anonymous referee for
his/her constructive comments and suggestions leading to
significant improvement of this paper. H. T. L. thanks for
financial support by National Natural Science Foundation of China
(NSFC, Grant 10533050 and 10778726). L. M. is supported by NSFC
(Grant 10778726). J. M. B. thanks support of the Bai Ren Ji Hua
project of the CAS, China and the 973 Program (Grant
2009CB824800).

\clearpage

\begin{figure}
\centering
\includegraphics[angle=-90,scale=0.5]{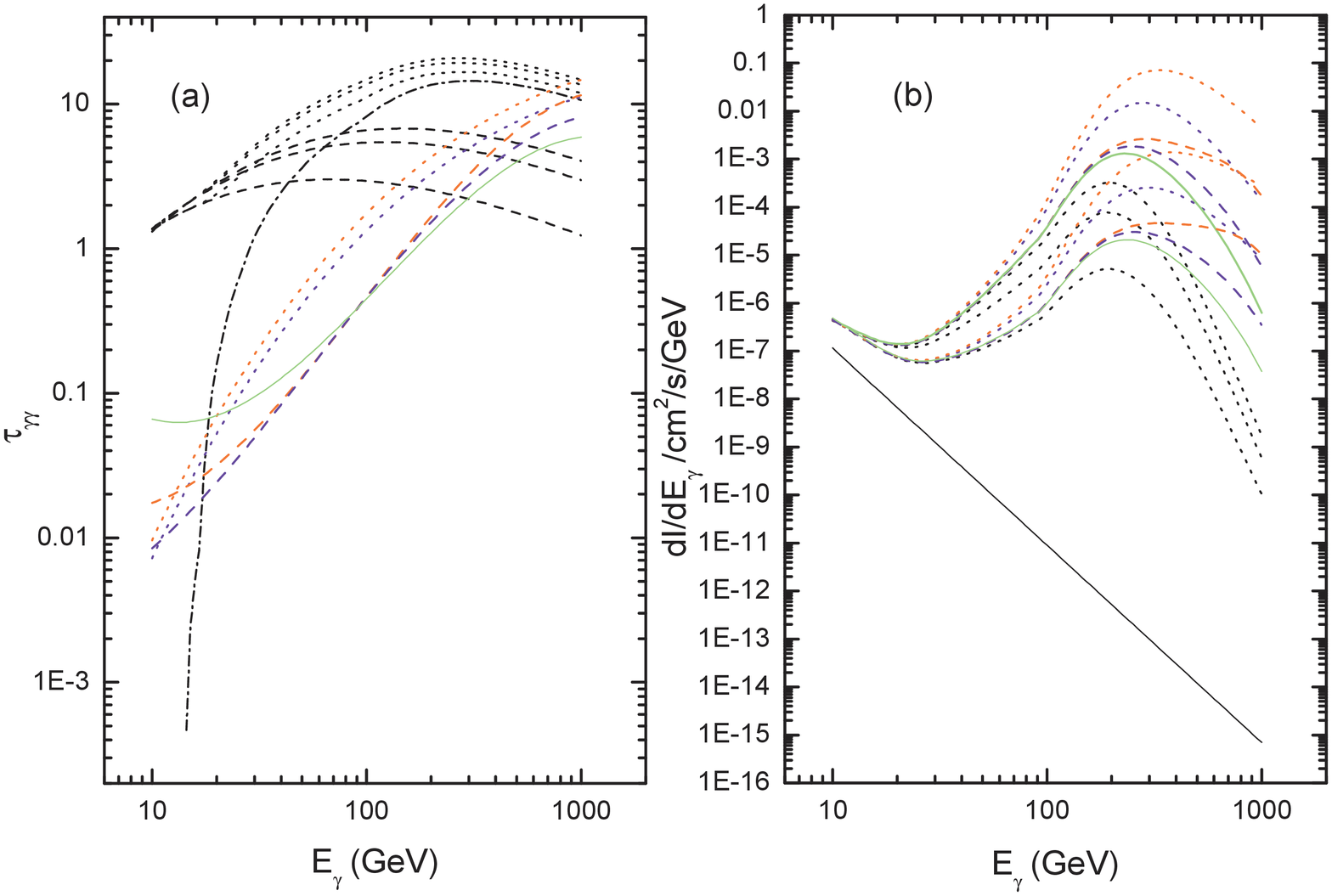}
 \caption{$(a)$ Photon-photon absorption optical depth for 3C 279. Dashed lines, absorption by
 multitemperature blackbody; dash-dotted line, absorption by broad emission lines; dotted lines,
 the total internal absorption by the two. In calculating $\tau_{\gamma\gamma}$, we assumed
 $R_{\rm{\gamma}}=r_{\rm{BLR,in}}$, $\tau_{\rm{BLR}}/f_{\rm{cov}}=1$, and $f_{\rm{cov}}=0.03$, and adopted
 $r_{\rm{BLR,in}}=0.1 \/\ \rm{pc}$, $r_{\rm{BLR,out}}=0.4 \/\ \rm{pc}$, $L_{\rm{BLR}}=10^{44.41}\/\
 \rm{ergs \/\ s^{-1}}$, and $M_{\rm{BH}}=10^{8.4} \/\ M_{\sun}$. From the top down, the dotted and dashed
 lines are for $a_{\rm{\ast}}=0.5$, $a_{\rm{\ast}}=0.8$, and $a_{\rm{\ast}}=0.998$. $(b)$ Gamma-ray
 spectrum of 3C 279: the spectrum given by eq.(1) (solid line) and the spectra corrected for the
 internal absorption (black dotted lines). From the top down, the black dotted lines are for $a_{\rm{\ast}}=0.998$,
  $a_{\rm{\ast}}=0.8$, and $a_{\rm{\ast}}=0.5$. Color lines in $(a)$ are the external absorption
  due to the EBL, estimated according to eq.(2): red and blue dotted lines are the fast evolution and baseline models,
  respectively; red and blue dashed lines the best-fit and low-IR Kneiske et al. models, respectively; green
  line the Primack et al. model. Color lines in $(b)$ correspond to the black dotted lines
  in $(b)$ for the cases of $a_{\rm{\ast}}=0.5$ and $a_{\rm{\ast}}=0.998$, but corrected for the EBL absorption
  described by the relevant color lines in ($a$).}
 \label{fig1}
\end{figure}

\begin{figure}
\centering
\includegraphics[angle=-90,scale=0.5]{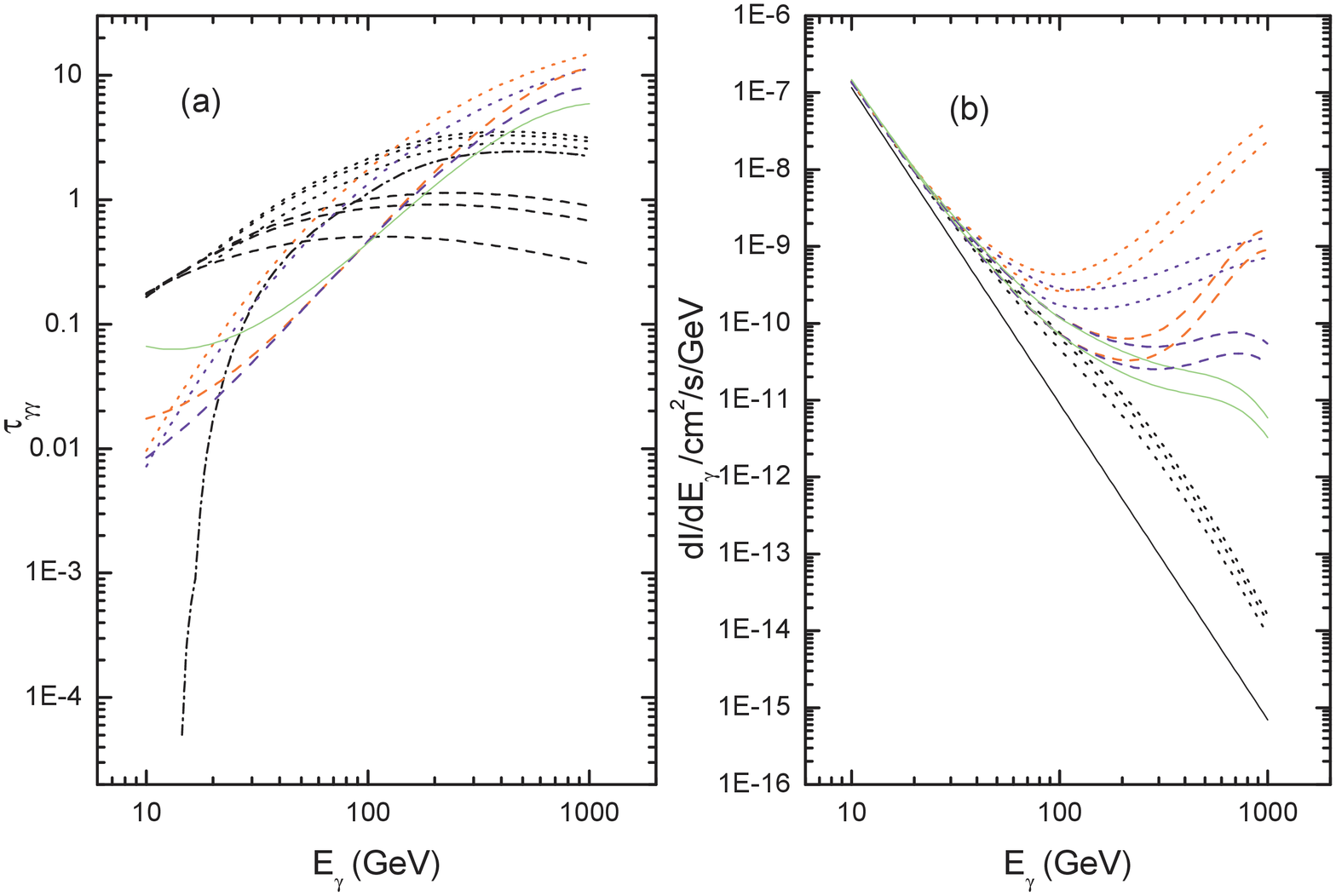}
 \caption{Same as Fig. 1, but for $R_{\rm{\gamma}}=(r_{\rm{BLR,in}}+r_{\rm{BLR,out}})/2=0.25 \/\
 \rm{pc}$.
  }
 \label{fig2}
\end{figure}

\begin{figure}
\centering
\includegraphics[angle=-90,scale=0.5]{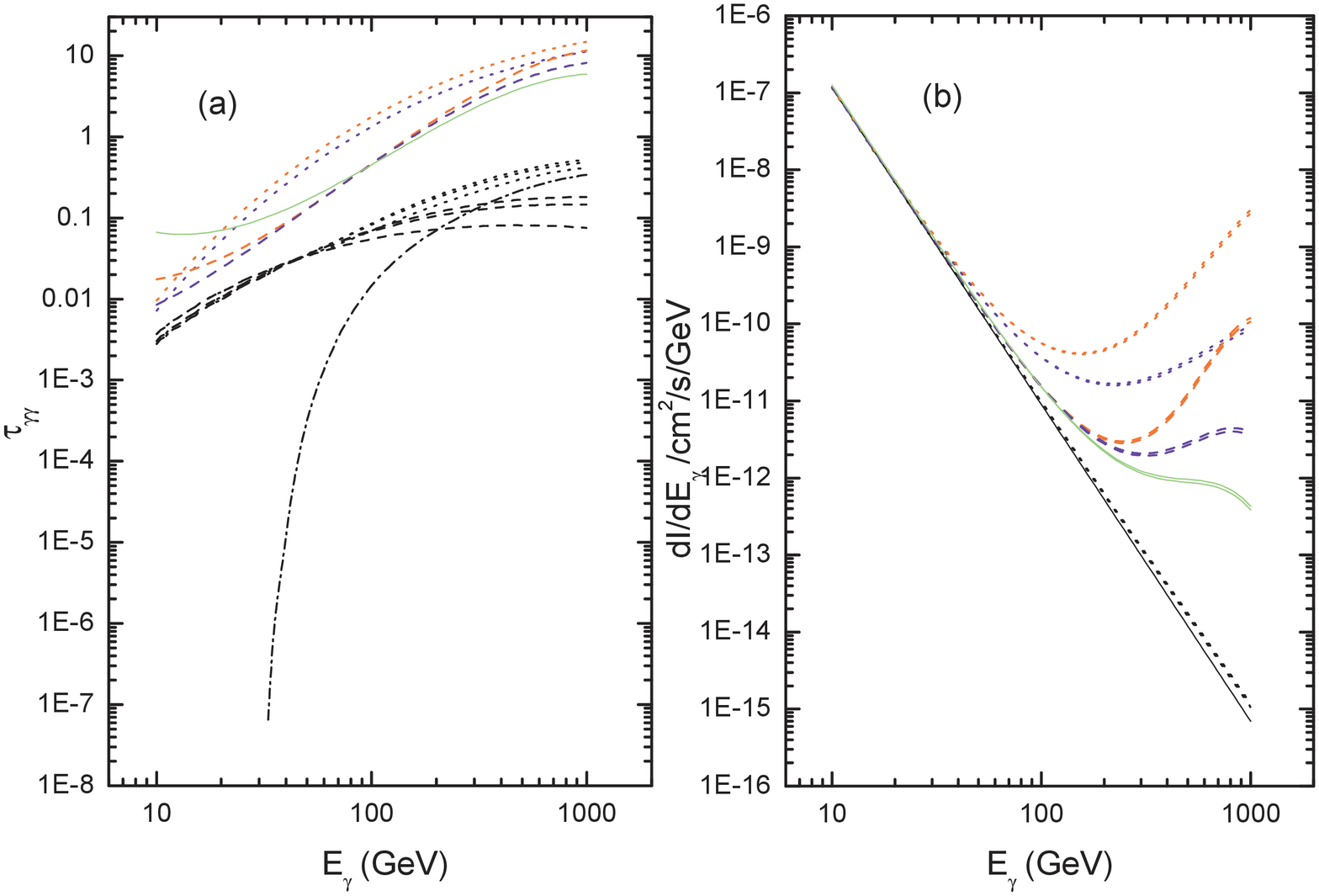}
 \caption{Same as Fig. 1, but for $R_{\rm{\gamma}}=r_{\rm{BLR,out}}$.
 }
 \label{fig3}
\end{figure}

\begin{figure}
\centering
\includegraphics[angle=-90,scale=0.5]{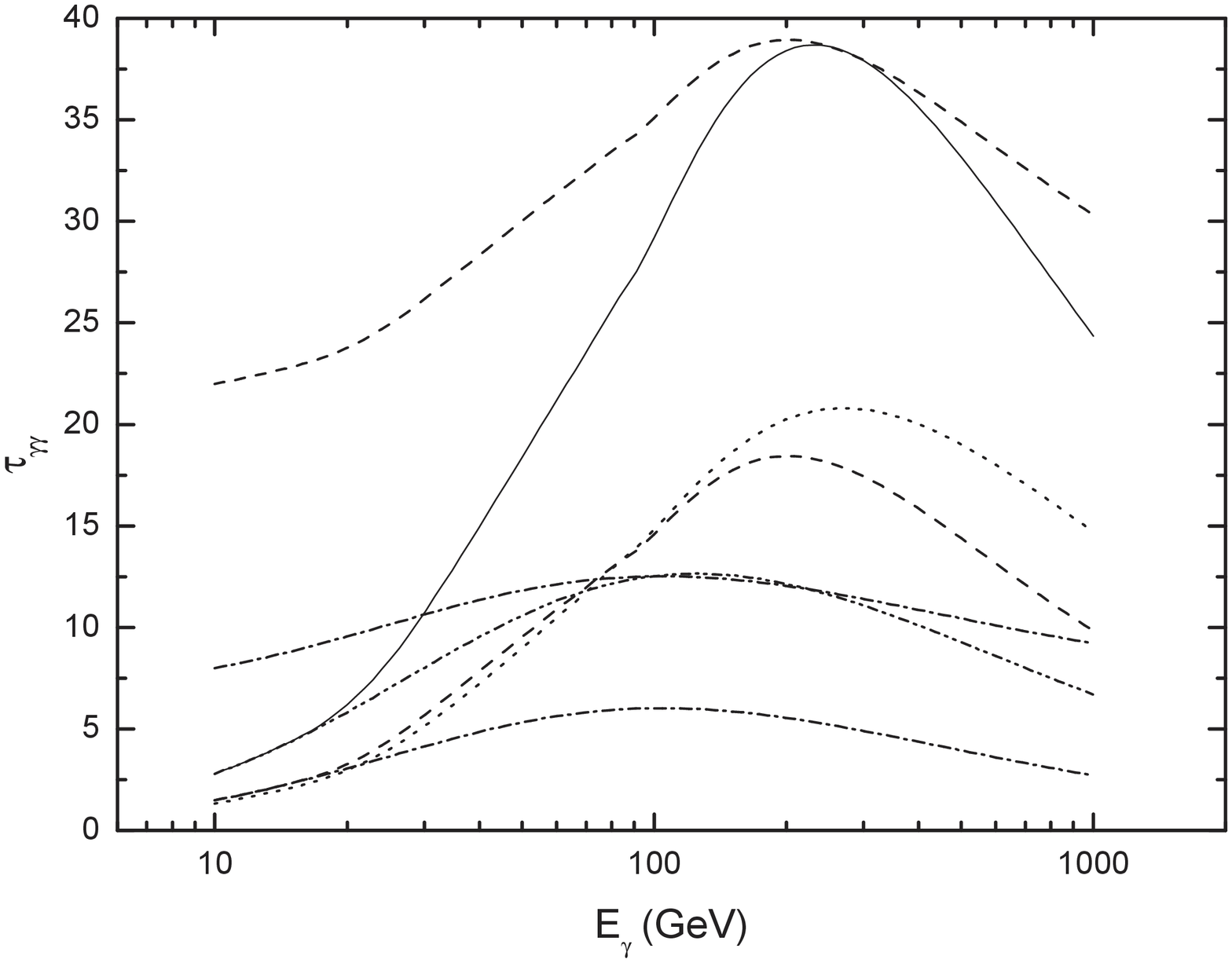}
 \caption{Photon-photon absorption optical depth for 3C 279 as $a_{\rm{\ast}}=0.5$, $r_{\rm{BLR,in}}=0.1 \/\ \rm{pc}$,
 $r_{\rm{BLR,out}}=0.4 \/\ \rm{pc}$, and $R_{\rm{\gamma}}=2\times 10^{16} \/\ \rm{cm}$. Solid line
 is the total absorption optical depth as $R_{1}=R_{\rm{\gamma}}$ and $R_{2}>>r_{\rm{BLR,in}}$. The lower dashed line
 is the total one as $R_{1}=R_{\rm{\gamma}}$ and $R_{2}=r_{\rm{BLR,in}}$. Dash-double-dotted line
 is the absorption optical depth due to the diffuse continuum as $R_{1}=R_{\rm{\gamma}}$ and $R_{2}>>r_{\rm{BLR,in}}$.
 The lower dash-dotted line is the absorption optical depth due to the diffuse continuum as $R_{1}=R_{\rm{\gamma}}$ and
 $R_{2}=r_{\rm{BLR,in}}$. The upper dashed line is the upward shifted one of the lower dashed line. The upper
 dash-dotted line is the upward shifted one of the lower dash-dotted
 line. Dotted line is the upper dotted line in plot $(a)$ in Fig. 1.
 }
 \label{fig4}
\end{figure}

\end{document}